\documentclass[twocolumn,trackchanges]{aastex631}
\hypersetup{linkcolor=red,citecolor=green,filecolor=cyan,urlcolor=magenta}

\usepackage{longtable}
\usepackage{booktabs}
\usepackage{float}
\usepackage{amsmath}
\shorttitle{RRLs in $\sl Kepler$/$\sl K2$-LAMOST Project}
\shortauthors{P. Zong et al.}
\graphicspath{{./}{figures/}}

\begin{document}

\title{A Study of Pulsation properties of 57 Non-Blazhko effect ab-type RR Lyrae stars with homogeneous metallicities from the LAMOST-$\sl Kepler$/$\sl K2$ survey}

\correspondingauthor{Jian-Ning, Fu}
\email{jnfu@bnu.edu.cn}

\author{Peng Zong}
\affiliation{Institute for Frontiers in Astronomy and Astrophysics, Beijing Normal University, Beijing 102206, People’s Republic of China}
\affiliation{Department of Astronomy, Beijing Normal University, Beijing 100875, People’s Republic of China}

\author{Jian-Ning Fu}
\affiliation{Institute for Frontiers in Astronomy and Astrophysics, Beijing Normal University, Beijing 102206, People’s Republic of China}
\affiliation{Department of Astronomy, Beijing Normal University, Beijing 100875, People’s Republic of China}

\author{Jiaxin Wang}
\affiliation{College of Science, Chongqing University of Posts and Telecommunications, Chongqing 400065, People’s Republic of China}

\author{Tian-Qi Cang}
\affiliation{Institute for Frontiers in Astronomy and Astrophysics, Beijing Normal University, Beijing 102206, People’s Republic of China}
\affiliation{Department of Astronomy, Beijing Normal University, Beijing 100875, People’s Republic of China}

\author{HaoTian Wang}
\affiliation{Institute for Frontiers in Astronomy and Astrophysics, Beijing Normal University, Beijing 102206, People’s Republic of China}
\affiliation{Department of Astronomy, Beijing Normal University, Beijing 100875, People’s Republic of China}

\author{Xiao-Yu Ma}
\affiliation{Institute for Frontiers in Astronomy and Astrophysics, Beijing Normal University, Beijing 102206, People’s Republic of China}
\affiliation{Department of Astronomy, Beijing Normal University, Beijing 100875, People’s Republic of China}

\author{Weikai Zong}
\affiliation{Institute for Frontiers in Astronomy and Astrophysics, Beijing Normal University, Beijing 102206, People’s Republic of China}
\affiliation{Department of Astronomy, Beijing Normal University, Beijing 100875, People’s Republic of China}



\begin{abstract}
Homogeneous metallicities and continuous high-precision light curves play key roles in studying the pulsation properties of RR Lyrae stars. By cross-matching with LAMOST DR6, we have determined 7 and 50 Non-Blazhko RRab stars in the {\sl Kepler} and {\sl K}2 fields, respectively, who have homogeneous metallicities determined from low-resolution spectra of the LAMOST-$\sl Kepler$/$\sl K2$ project. The Fourier Decomposition method is applied to the light curves of these stars provided by the $\sl Kepler$ space based telescope to determine the fundamental pulsation periods and the pulsation parameters. The calculated amplitude ratios of $R_{21}$, $R_{31}$ and the phase differences of $\phi_{21}$, $\phi_{31}$ are consistent with the parameters of the RRab stars in both the Globular Clusters and the Large Magellanic Cloud. We find a linear relationship between the phase differences $\phi_{21}$ and $\phi_{31}$, which is in good agreement with the results in previous literature. As far as the amplitude, we find that the amplitude of primary frequency A$_1$ and the total amplitude A$_{tot}$ follow either a cubic or linear relationship. For the rise time $RT$, we do not find its relevance with the period of the fundamental pulsation mode P$_1$, or A$_{tot}$ and $\phi_{21}$.  However, it might follow a linear relationship with R$_{31}$. Based on the homogeneous metallicities, we have derived a new calibration formula for the relationship of period-$\phi_{31}$-[Fe/H], which agrees well with the previous studies.
\end{abstract}



\section{Introduction} \label{sec:intro}
RR Lyrae stars (hereafter RRLs) are low-mass pulsating stars located at the intersection between the classical instability strip and the horizontal branch of the Hertzsprung-Russell diagram, with helium burning in the core \citep{2010aste.book.....A}.  
Their pulsations are induced by the so-called  $\kappa$-mechanism, operating in the hydrogen and helium partial ionization zones. Typical RRLs have pulsation periods of $0.2-1$\,d, amplitudes of $0.3^{\mathrm{m}}-1^{\mathrm{m}}$, effective temperature $T_\mathrm{eff}$ of 6100-7400~K, and spectral types of A2-F6 \citep{2015pust.book.....C}. According to the pulsation modes, they can be divided into three types, i.e., the foundamental radial mode (type RRab), the first overtone (type RRc) or both modes simultaneously (type RRd) \citep{2012MNRAS.424..649G,2015MNRAS.447.2348M}.

RRLs are widely used as tracers of stellar populations with ages older than 10 Gyr in the Milky Way \citep{1989PASP..101..570W, 2021ApJ...912..144M} and neighboring galaxies \citep{2015pust.book.....C, 2021FrASS...7...81P}. They are also commonly used as standard candles, benefiting from their relationship between M$_V$ and iron abundance \citep{1993AJ....106..703S,1998A&ARv...9...33C,2013ApJ...773..181N}. However, this relationship's intrinsic and systematic errors are still in debate \citep[see, e.g.,][]{2000MNRAS.316..819C,2004ApJ...612.1092D,2008ApJ...672L.115C,2012MSAIS..19..138M,2009AIPC.1170..223M,2021ApJ...912..144M}. The metallicities of RRLs in those studies have been highlighted for a reliable estimation of distance with the relation of period-luminosity-metallicity. But it is difficult to obtain accurate measurements of metal abundances of this type of stars \citep{2011ApJS..194...38F,2014ApJ...782...59G,2013ApJ...773..181N,2017ApJ...848...68S,2017ApJ...835..187C,2019ApJ...881..104M,2021ApJ...908...20C,2021MNRAS.503.4719G}. \cite{2022arXiv220401953A} reported a homogeneous approach towards the calculation of mean M$_V$ and [Fe/H] with a sample of 37 globular clusters via the Fourier decomposition of their light curves. Another interesting astrophysical problem of RRLs is the Blazhko effect, which is the amplitude and phase modulation of the light curves on the time-scale of tens to thousands of days \citep{1907AN....175..325B,1916ApJ....43..217S}. Almost 30$\%$-50$\%$ RRLs exhibit Blazhko characteristics, but the physical origin of the effect is still a mystery since its discovery \citep{2014ApJS..213...31B}. 

The investigations of RRLs have made great progress with the unprecedented and ever-precious photometric data obtained by the $\sl Kepler$ and $\sl K2$ mission. \cite{2010MNRAS.409.1585B} investigated a sample of 29 RRLs using the $\sl Kepler$ photometry and found that almost half of the sample exhibited Blazhko effect. The prototype star RR Lyr itself was studied in detail with the Q1-Q2 long cadence (LC) data of $\sl Kepler$ by \cite{2011MNRAS.411..878K}, who found a multiplet structure at the main frequency and its harmonics up to the quintuplets. \cite{2011MNRAS.417.1022N} carried out Fourier analysis of 19 non-Blazhko RRab stars with {\sl Kepler} photometry, among which none of the stars showed the period-doubling effect as seen in Blazhko stars. They also found that KIC 7021124 pulsates simultaneously in both the fundamental and second overtone modes. Based on the following high-resolution spectroscopic observations with CFHT and Keck-I, \citet{2013ApJ...773..181N} determined the iron-to-hydrogen ratios, radial velocities, and atmospheric parameters for 41 RRLs in the $\sl Kepler$ field and thus gave a new relationship of Period-$\phi_{31}$-[Fe/H]. \cite{2022AJ....164...45N} adopted a set of homogeneous samples of fundamental mode RRLs in $\sl Kepler$ field to investigate the performance of photometric metallicity. Comparing with roughly 50 RRLs in the prime $\sl Kepler$ field, more than 3000 RRLs had been  proposed for observations in $\sl K2$ campaigns. Although one might lose the chance of studying Blazhhko effect of RRLs merely with $\sl K2$ photometry concerning the limited lengths of the time-series observations of the target stars, it is possible to carry out population studies \citep{2015MNRAS.452.4283M, 2016MNRAS.456.2260A} and statistical investigations \citep{2018A&A...614L...4K,2021arXiv211009279M}. 

\cite{1996A&A...312..111J} have revealed that the shapes of light curves of RRab stars in optical wavelength have relevance with their mental abundances. They derived a linear relation of mental abundance with period and low-order parameters of Fourier Decomposition on the light curves of RRab stars in V band, with the phase difference of $\phi_{31}$ = $\phi_{3}$-3$\phi_{1}$. 
This relation was investigated by \cite{2005AcA....55...59S} with the light curves in I band of RRab stars from the Optical Gravitational Lensing Experiment \citep{1992AcA....42..253U}. A calibration of the relation was carried out by \cite{2016ApJS..227...30N} using the data of the Palomar Transient Factory \citep{2009PASP..121.1395L} in R band. 
\cite{2013ApJ...773..181N} extended the analysis of this relation using well-sampled light curves of RRab stars in the $\sl Kepler$ field \citep{2010ApJ...713L..79K}. The calibration of the relation was given by \cite{2016CoKon.105...53M} using the RRab stars in globular clusters (GCs) and fields. \cite{2021MNRAS.502.5686I} obtained a new Period-$\phi_{31}$-[Fe/H] relation using G-band light curves provided by Gaia DR2 \citep{2018A&A...616A..12G,2018A&A...618A..30H,2019A&A...622A..60C}. Recent study of \cite{2021ApJ...912..144M} not only gave new relations adopting the light curves of RRab stars of ASAS-SN sample in V band, but also provided new calibrations for the stars observed by W1 and W2 WISE bands. A similar study for RRc stars was also carried out by \cite{2022ApJ...931..131M}. \cite{2022MNRAS.517.1907J} studied RRab stars with quasi-identical-shape light curves but period differences as large as 0.05-0.21 d based on the Galactic bulge data of the OGLE-IV survey. They revealed that several of these stars show very similar light curves to that of the typical bulge RR Lyrae by examining their Fourier parameters. However, to precisely characterize the relation of Period-$\phi_{31}$-[Fe/H] and the connection of pulsation parameters of RRab stars, homogeneous spectra and light curves are required to derive the metallicity abundances and pulsation parameters, respectively. Fortunately, observations of the LAMOST-$\sl Kepler$/$\sl K2$ project (LKS) \citep[see,e.g.,][]{2015ApJS..220...19D,2018ApJS..238...30Z,2020ApJS..251...27W,2020RAA....20..167F} have provided LAMOST spectra for a larger number of $\sl Kepler$/$\sl K2$ targets. In this study, we investigate the characteristics of the pulsation parameters of the non-Blazhko RRab stars based on the $\sl Kepler$ light curves and homogeneous metal abundances provided by the LAMOST-$\sl Kepler$/$\sl K2$ project, and a new calibration of Period-$\phi_{31}$-[Fe/H] is presented.

This paper is organized as following: the target selection process is described in $\S$ \ref{sec:style}. The Fourier decomposition analysis of light curves is presented in $\S$ 3. The analysis results and discussion are given in $\S$ 4 and $\S$ 5, respectively. Finally, we present conclusions of this work in $\S$ 6.

\section{Target selection} \label{sec:style}
\begin{figure*}
\includegraphics[width=7.0 in]{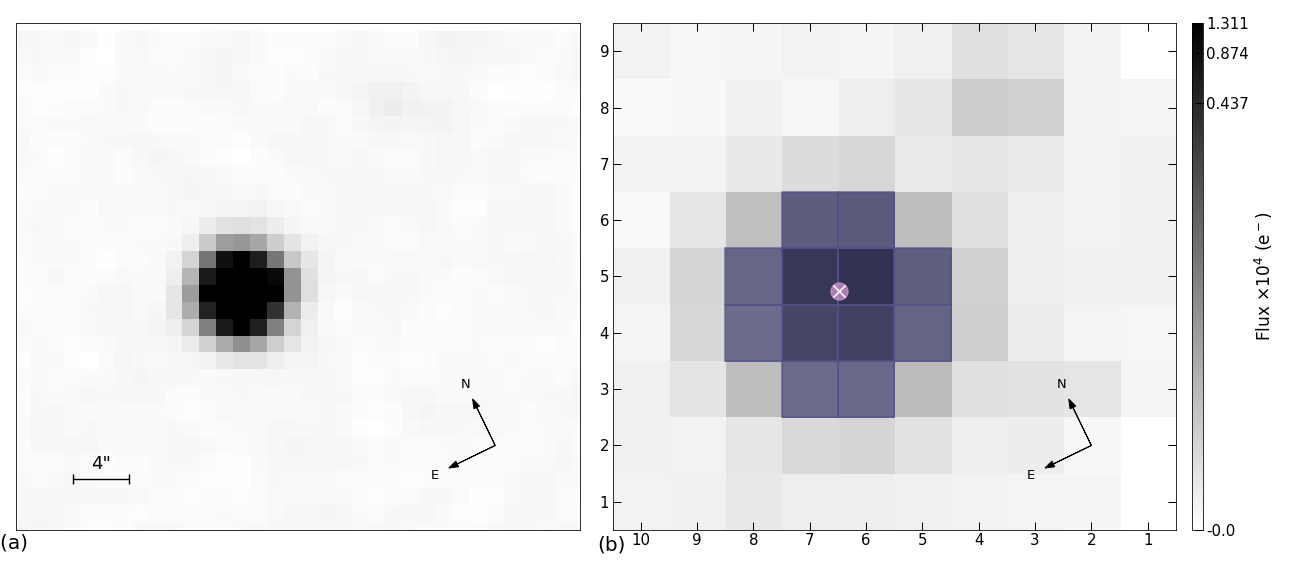}
\caption{The image of the non-Blazhko RRab star EPIC 210830646. (a) The target pixel file of the star; (b) the green polygon indicates the optimized aperture adopted on the star for photometry.}
\label{fig:TPF}
\end{figure*}

\begin{figure*}
\includegraphics[width=7.0 in]{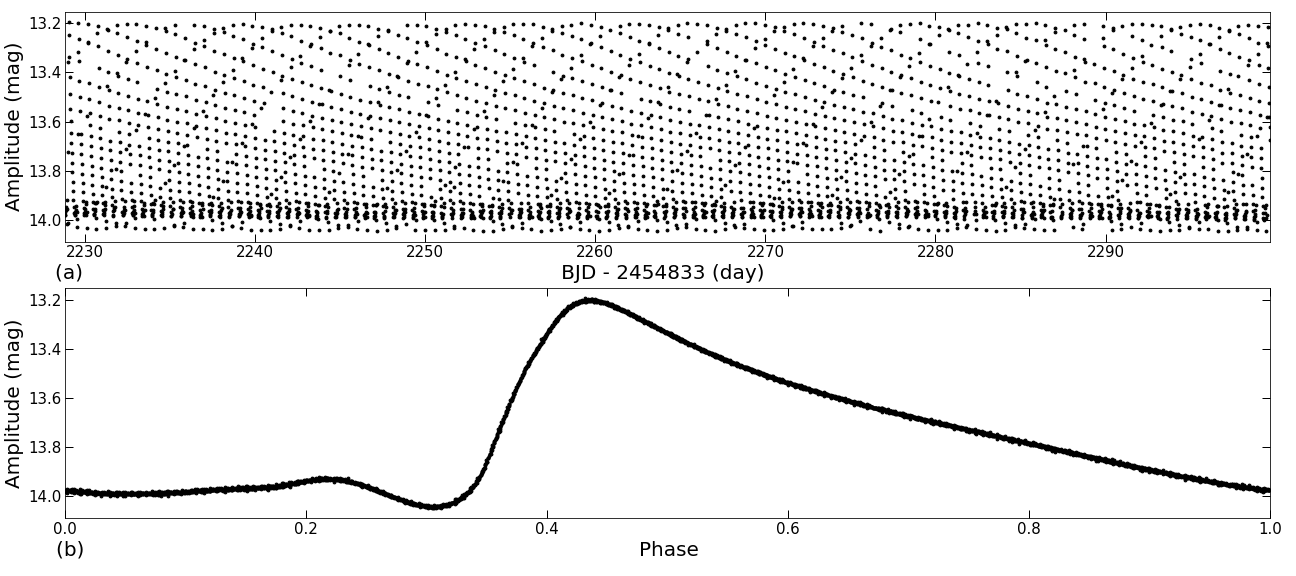}
\caption{ (a) The light curve of EPIC 210830646 extracted by LightKurve \citep{2018zndo...1181929V,2020zndo...1181928B}, (b) the phase-folded light curve in fundamental period.}
\label{fig:LC}
\end{figure*}

The catalog of RRLs in the fields of $K2$ is obtained by combining the catalogs of RRLs of all campaigns downloaded from EVEREST \citep{2016AJ....152..100L,2018AJ....156...99L}. We obtain the target pixel files (hereafter TPFs) of all candidates of 3413 stars at the Mikulski Archive for Space Telescopes \footnote{\url{https://archive.stsci.edu/missions-and-data/kepler}} (MAST, all the {\it K2} data used in this paper can be found in MAST: \dataset[10.17909/T9K30X]{http://dx.doi.org/10.17909/T9K30X})  using the catalogs. The light curves are then extracted from TPFs with the LightKurve package \citep{2018zndo...1181929V,2020zndo...1181928B}. For each star, a series of apertures are tested on the TPFs in order to optimize the photometry precision. The extracted light curves are then detrended by fitting and subtracting either a second- or third-order polynomial to remove the long-term systematic errors. Finally, The corresponding fluxes are converted to magnitudes and shifted to the $K{\rm p}$ mean magnitude levels. As an example, the images and light curves of the non-Blazhko RRab star EPIC 210830646 observed by $\sl K2$ are shown in Figures~\ref{fig:TPF} and \ref{fig:LC}, respectively.

We identify the non-Blazhko and Blazhko RRLs among those candidates by following the most strict and convincing evidence whether the presence of the side peaks is shown in the frequency spectra \citep{2016A&A...592A.144S}. We searched for this feature with the software Period04 \citep{2005CoAst.146...53L}, and identified 376 Blazhko and 594 non-Blazhko RRab stars, respectively. Then, we take those non-Blazhko RRab stars to cross-match with the catalog of \cite{2020ApJS..247...68L}, who derived the metal abundances of RRLs from the low-resolution spectra of LAMOST DR6, with a total of 50 non-Blazhko RRab stars matched. We also cross-match the catalog of \cite{2020ApJS..247...68L} with a list of the $\sl Kepler$ non-Blazhko RRab stars in previous study \citep{2013ApJ...773..181N}, among which 7 stars are derived. As \cite{2020ApJS..247...68L} did not give the uncertainties of the metal abundances of the stars, we estimate the uncertainties using the method provided by \cite{2020ApJS..251...27W}. The values of the metal abundances of the 57 non-Blazhko RRab stars, with their corresponding uncertainties of the metal abundances, are listed in the sixth column of Table~\ref{tab:parameters}.


\section{Pulsation analysis}
Frequency analysis with Fourier Decomposition is useful to characterize pulsations of RRLs (e.g., \citealt{1982ApJ...261..586S,1993AJ....106..703S}). 
For the non-Blazhko RRab stars from No.8 to No.57 in Table~\ref{tab:parameters}, frequency analyses of the light curves are carried out with the software Period04. The corresponding uncertainties of the frequencies and periods are determined according to the method proposed by \cite{2021ApJ...921...37Z}. After the frequencies are extracted from the Fourier amplitude spectra, the light curves are fitted with the following formula concerning the sine function series,

\begin{equation}
\label{eq:1}
m(t) = m_0 + \sum_{i=1}^{n} A_i{\rm sin}(2\pi i f_0(t-t_0)+\phi_i)
\end{equation}

where $n$ is the number of fitted orders, $f_0$ the main frequency, $t$ the observation time (Barycentric Julian Date: BJD-2454833.0 ) and $t_0$ the time of the first minimum apparent magnitude of the light curves. The mean magnitude $m_0$, amplitude A$_i$, and phase $\phi_{i}$ values at given $i$th-order can be determined. As \cite{1981ApJ...248..291S} suggested, certain combinations of Fourier coefficients are directly related to some physical parameters of pulsating stars. These coefficients are typically defined either as linear combinations of phase difference of $\phi_{ij}$ or as amplitude ratios $R_{ij}$ as follows,
\begin{equation}
\label{eq:2}
\phi_{ij} = j\phi_{i}-i\phi_{j}
\end{equation}

\begin{equation}
\label{eq:3}
 R_{ij} = \frac{A_i}{A_j}
\end{equation}
where i = 2 or 3, j = 1 for the fundamental modes of RRLs as suggested by \cite{2013ApJ...773..181N} and \cite{2013MNRAS.428.3034S}. Note that $\phi_{21}$ and $\phi_{31}$ are corrected for integer multiples of $\pi$ to meet the $\phi_{21}$ $<$ $\pi$ and $\pi$ $<$ $\phi_{31}$ $<$ 2$\pi$ conditions. The parameters of 5 $\sl Kepler$ RRab stars analyzed from No.1 to No.4 and No.7 of Table~\ref{tab:parameters} are taken from \cite{2011MNRAS.417.1022N}. We also calculate the maximum (A$_{max}$) and minimum (A$_{min}$) light with their corresponding phases ($\phi_{max}$ and $\phi_{min}$) of those stars in $\sl Kepler$ and $ \sl K2$ missions, by fitting a second or third degree polynomial around each peak and valley of the phase-folded light curves, which can be used to determine the rise time RT = $\phi_{max}$ - $\phi_{min}$ and total amplitude A$_{\rm tot}$ = A$_{max}$ - A$_{min}$ for each star. The parameters of the stars KIC 9658012 and 9717032 as No.5 and 6 in Table~\ref{tab:parameters} are determined in this work. The uncertainties of the Fourier coefficients for the 57 stars are also estimated. Table~\ref{tab:parameters} lists the properties of these non-Blazhko RRLs in the fields of LAMOST-$Kepler$/$K2$ project.

\renewcommand\tabcolsep{2pt}
\startlongtable
\begin{longrotatetable}
\begin{deluxetable}{lcccccccccccc}
\tablecaption{Properties of non-Blazhko RRab stars in the field of LAMOST-$Kepler$/$K2$ survey. No. of star (Column 1), ID of the target star (Column 2), $\sl K_p$ magnitude (Column 3), pulsation frequency and corresponding period ( Column 4-5), metal abundance (Column 6), the Fourier A$_1$ coefficient (Column 7), total amplitude (Column 8), rise time (RT) (Column 9), and Fourier-based coefficient (Column 10-13).\label{tab:parameters}}
\tablehead{No.&ID       & $\sl Kp$ & Frequency &Period & [Fe/H]  &  A$_1$ & RT &  A$_{tot}$ & R$_{21}$ & R$_{31}$ & $\phi_{21}$ & $\phi_{31}$ \\
& KIC/EPIC  &(mag)   & (c/d)  &  (day)   & (dex)     &   (mag)  & (rad) &  (mag)      &          &          &    (rad)      &      (rad)  \\
(1)& (2)    &  (3)     &  (4) &(5)     & (6)       & (7)    &  (8)   &  (9)         & (10)     &    (11)  &  (12)    & (13)  \\}

\startdata
 1&6100702 & 13.458&2.048567(8) &0.488146(2)&-0.60\textbf{(6)} & 0.2089\textbf{(5)} &0.200\textbf{(4)} &0.575\textbf{(9)} &0.493\textbf{(2)} &0.279\textbf{(2)} &2.743\textbf{(1)} &5.747\textbf{(1)} \\
 2&6763132 & 13.075&1.701291(3) &0.587789(1) &-1.90\textbf{(7)} &0.2802\textbf{(11)} &0.144\textbf{(3)} &0.811\textbf{(19)} &0.471\textbf{(4)} &0.356\textbf{(11)} &2.389\textbf{(2)} &5.096\textbf{(4)} \\
 3&7021124 & 13.550&1.606443(2) &0.622493(7) &-1.65\textbf{(7)} &0.0323\textbf{(2)} &0.139\textbf{(1)} &0.831\textbf{(5)} &0.512\textbf{(5)} &0.351\textbf{(5)} &2.372\textbf{(2)} &5.060\textbf{(2)} \\ 
 4&7030715 & 13.452&1.462813(4) &0.683614(2) &-1.45\textbf{(7)} &0.2075\textbf{(6)} &0.177\textbf{(3)} &0.647\textbf{(15)} &0.494\textbf{(3)} &0.303\textbf{(2)} &2.683\textbf{(1)} &5.606\textbf{(1)} \\
 5&9658012 & 16.001&1.875482(5) &0.533196(3) &-0.70\textbf{(6)} &0.3108\textbf{(14)} &0.141\textbf{(3)} &0.928\textbf{(30)} &0.498\textbf{(5)} &0.360\textbf{(4)} &2.399\textbf{(2)} &5.156\textbf{(3)} \\
 6&9717032 & 17.194&1.795615(9) &0.556912(5) &-1.40\textbf{(6)} &0.2119\textbf{(11)} &0.149\textbf{(5)} &0.633\textbf{(21)} &0.512\textbf{(5)} &0.353\textbf{(5)} &2.403\textbf{(2)} &5.139\textbf{(2)} \\
 7&9947026 & 13.300&1.822858(3) &0.548589(1) &-2.60\textbf{(7)} &0.2170\textbf{(5)} &0.193\textbf{(12)} &0.599\textbf{(27)} &0.488\textbf{(2)} &0.282\textbf{(2)} &2.745\textbf{(1)} &5.737\textbf{(2)} \\
 8&201395405 & 17.558&1.622410(1) &0.616367(7) &-1.55\textbf{(6)} &0.1661\textbf{(12)} &0.160\textbf{(3)} &0.412\textbf{(10)} &0.386\textbf{(6)} &0.320\textbf{(6)} &2.380\textbf{(17)} &4.92\textbf{(31)} \\ 
 9&201416966 & 14.441& 1.767199(7)&0.565867(4) &-2.10\textbf{(6)} &0.2719\textbf{(13)} &0.156\textbf{(3)} &0.804\textbf{(15)} &0.506\textbf{(6)} &0.346\textbf{(5)} &2.466\textbf{(14)} &5.27\textbf{(30)} \\
 10&201454019 & 14.560&1.41163(1) &0.708401(7) &-1.30\textbf{(6)} &0.1637\textbf{(7)} &0.188\textbf{(3)} &0.416\textbf{(9)} &0.497\textbf{(4)} &0.305\textbf{(3)} &2.587\textbf{(9)} &5.91\textbf{(24)} \\
 11&201458397 & 16.842&1.604129(1) &0.623391(6) &-1.50\textbf{(7)} &0.1884\textbf{(5)} &0.221\textbf{(5)} &0.494\textbf{(8)} &0.442\textbf{(3)} &0.253\textbf{(3)} &2.717\textbf{(9)} &5.71\textbf{(23)} \\ 
 12&201550621 & 12.626&1.841698(6) &0.542977(3) &-1.50\textbf{(6)} &0.2995\textbf{(16)} &0.159\textbf{(7)} &0.893\textbf{(14)} &0.523\textbf{(6)} &0.345\textbf{(6)} &2.472\textbf{(15)} &5.28\textbf{(31)} \\ 
 13&201611200 & 14.459&1.699969(10) &0.588246(6) &-0.95\textbf{(6)} &0.1703\textbf{(8)} &0.173\textbf{(3)} &0.465\textbf{(11)} &0.500\textbf{(4)} &0.329\textbf{(4)} &2.590\textbf{(10)} &5.50\textbf{(26)} \\ 
 14&202064491 & 15.000&1.746749(2) &0.572492(1) &-1.20\textbf{(6)} &0.1959\textbf{(15)} &0.161\textbf{(4)} &0.534\textbf{(14)} &0.500\textbf{(7)} &0.344\textbf{(7)} &2.496\textbf{(17)} &5.34\textbf{(33)} \\ 
 15&210345648 & 15.087&1.686611(9) &0.592905(6) &-1.20\textbf{(3)} &0.1960\textbf{(12)} &0.149\textbf{(5)} &0.583\textbf{(19)} &0.422\textbf{(6)} &0.347\textbf{(5)} &2.401\textbf{(16)} &5.01\textbf{(31)} \\ 
 16&210454384 & 17.525&1.690231(9) &0.591635(5) &-1.70\textbf{(6)} &0.2128\textbf{(11)} &0.186\textbf{(7)} &0.609\textbf{(16)} &0.500\textbf{(6)} &0.314\textbf{(5)} &2.639\textbf{(15)} &5.56\textbf{(31)} \\ 
 17&210766289 & 16.398&1.774371(9) &0.563580(5) &-1.25\textbf{(6)} &0.2085\textbf{(13)} &0.148\textbf{(3)} &0.619\textbf{(13)} &0.489\textbf{(7)} &0.357\textbf{(7)} &2.036\textbf{(19)} &5.57\textbf{(35)} \\ 
 18&210815465 & 16.002&1.918090(5) &0.521352(2) &-1.60\textbf{(6)} &0.3826\textbf{(22)} &0.137\textbf{(5)} &1.140\textbf{(12)} &0.467\textbf{(6)} &0.372\textbf{(6)} &2.344\textbf{(17)} &4.91\textbf{(32)} \\ 
 19&210830646 & 13.751&1.758810(8) &0.568566(5) &-1.70\textbf{(5)} &0.2328\textbf{(16)} &0.129\textbf{(1)} &0.628\textbf{(19)} &0.415\textbf{(6)} &0.336\textbf{(6)} &2.356\textbf{(17)} &4.80\textbf{(32)} \\ 
 20&210862813 & 16.312&1.441879(7) &0.693539(5) &-0.85\textbf{(6)} &0.2223\textbf{(3)} &0.214\textbf{(6)} &0.634\textbf{(5)} &0.481\textbf{(5)} &0.263\textbf{(4)} &2.697\textbf{(12)} &5.62\textbf{(28)} \\ 
 21&210872065 & 16.016&1.593511(7) &0.627545(4) &-1.10\textbf{(6)} &0.2369\textbf{(5)} &0.191\textbf{(5)} &0.637\textbf{(5)} &0.457\textbf{(4)} &0.289\textbf{(3)} &2.653\textbf{(10)} &5.63\textbf{(24)} \\ 
 22&210934522 & 15.477&2.297878(5) &0.435184(2) &-1.40\textbf{(6)} &0.3226\textbf{(16)} &0.147\textbf{(6)} &0.972\textbf{(25)} &0.533\textbf{(6)} &0.368\textbf{(6)} &2.346\textbf{(14)} &5.04\textbf{(31)} \\ 
 23&211410664 & 14.280&1.830881(8) &0.546185(4) &-1.15\textbf{(5)} &0.2275\textbf{(14)} &0.149\textbf{(3)} &0.654\textbf{(21)} &0.507\textbf{(6)} &0.354\textbf{(5)} &2.417\textbf{(14)} &5.19\textbf{(30)} \\ 
 24&211449132 & 15.008&1.772059(7) &0.564315(4) &-1.70\textbf{(6)} &0.2384\textbf{(12)} &0.148\textbf{(2)} &0.666\textbf{(11)} &0.476\textbf{(5)} &0.359\textbf{(5)} &2.389\textbf{(14)} &5.12\textbf{(29)} \\
 25&211484212 & 16.047&1.784809(7) &0.560284(4) &-1.45\textbf{(7)} &0.2388\textbf{(13)} &0.152\textbf{(4)} &0.693\textbf{(12)} &0.513\textbf{(5)} &0.347\textbf{(5)} &2.465\textbf{(13)} &5.27\textbf{(29)} \\ 
 26&211517414 & 15.601&1.537449(7) &0.650428(4) &-1.70\textbf{(6)} &0.2475\textbf{(8)} &0.183\textbf{(6)} &0.704\textbf{(14)} &0.510\textbf{(4)} &0.314\textbf{(3)} &2.717\textbf{(9)} &5.65\textbf{(24)} \\ 
 27&211580649 & 14.776&1.960661(7) &0.510032(4) &-1.35\textbf{(6)} &0.2491\textbf{(18)} &0.130\textbf{(4)} &0.705\textbf{(25)} &0.460\textbf{(6)} &0.358\textbf{(6)} &2.320\textbf{(16)} &4.88\textbf{(32)} \\ 
 28&211620065 & 16.395&2.034389(7) &0.491548(3) &-1.15\textbf{(7)} &0.2463\textbf{(17)} &0.136\textbf{(3)} &0.715\textbf{(21)} &0.474\textbf{(6)} &0.363\textbf{(6)} &2.313\textbf{(16)} &4.89\textbf{(31)} \\ 
 29&211653001 & 13.329&1.906039(7) &0.524648(3) &-2.00\textbf{(6)} &0.2478\textbf{(17)} &0.132\textbf{(3)} &0.715\textbf{(23)} &0.449\textbf{(6)} &0.353\textbf{(6)} &2.319\textbf{(16)} &4.84\textbf{(31)} \\ 
 30&211663101 & 14.365&1.665351(5) &0.600474(3) &-0.50\textbf{(6)} &0.3256\textbf{(16)} &0.150\textbf{(6)} &0.952\textbf{(19)} &0.493\textbf{(5)} &0.346\textbf{(5)} &2.418\textbf{(14)} &5.16\textbf{(30)} \\ 
 31&211918301 & 16.075&1.643631(6) &0.608409(3) &-1.35\textbf{(6)} &0.2831\textbf{(6)} &0.188\textbf{(8)} &0.722\textbf{(17)} &0.454\textbf{(3)} &0.293\textbf{(3)} &2.576\textbf{(9)} &5.50\textbf{(24)} \\ 
 32&211933496 & 15.414&2.058359(7) &0.485824(3) &-1.75\textbf{(6)} &0.2516\textbf{(19)} &0.132\textbf{(1)} &0.743\textbf{(19)} &0.457\textbf{(6)} &0.353\textbf{(6)} &2.314\textbf{(16)} &4.85\textbf{(31)} \\ 
 33&212130982 & 13.501&1.673371(7) &0.597596(4) &-1.20\textbf{(6)} &0.2614\textbf{(7)} &0.178\textbf{(2)} &0.772\textbf{(16)} &0.493\textbf{(4)} &0.311\textbf{(4)} &2.600\textbf{(10)} &5.51\textbf{(25)} \\
 34&212182292 & 12.392&1.832240(6) &0.545780(3) &-0.45\textbf{(6)} &0.2679\textbf{(9)} &0.172\textbf{(5)} &0.792\textbf{(13)} &0.513\textbf{(4)} &0.326\textbf{(3)} &2.757\textbf{(9)} &5.72\textbf{(24)} \\ 
 35&212789652 & 13.764&1.382640(6) &0.723254(4) &-1.20\textbf{(6)} &0.2701\textbf{(3)} &0.247\textbf{(2)} &0.794\textbf{(8)} &0.406\textbf{(2)} &0.190\textbf{(2)} &2.940\textbf{(6)} &6.15\textbf{(19)} \\ 
 36&212808200 & 13.259&1.534780(5) &0.651559(3) &-2.20\textbf{(6)} &0.3152\textbf{(15)} &0.159\textbf{(3)} &0.928\textbf{(23)} &0.513\textbf{(5)} &0.340\textbf{(5)} &2.489\textbf{(14)} &5.27\textbf{(30)} \\ 
 37&212870977 & 14.714&1.971449(7) &0.507241(3) &-1.15\textbf{(6)} &0.2713\textbf{(19)} &0.140\textbf{(8)} &0.798\textbf{(19)} &0.507\textbf{(6)} &0.358\textbf{(6)} &2.375\textbf{(16)} &5.10\textbf{(32)} \\ 
 38&220277833 & 13.269&2.214118(6) &0.451647(3) &-0.90\textbf{(7)} &0.2998\textbf{(2)} &0.134\textbf{(3)} &0.839\textbf{(22)} &0.470\textbf{(6)} &0.347\textbf{(5)} &2.290\textbf{(15)} &4.83\textbf{(31)} \\ 
 39&220288040 & 14.083&1.842869(6) &0.542632(3) &-1.40\textbf{(6)} &0.2911\textbf{(13)} &0.154\textbf{(6)} &0.861\textbf{(25)} &0.487\textbf{(5)} &0.357\textbf{(5)} &2.384\textbf{(14)} &5.12\textbf{(29)} \\ 
 40&220435063 & 12.783&1.484671(6) &0.673550(3) &-2.95\textbf{(6)} &0.2844\textbf{(14)} &0.146\textbf{(2)} &0.815\textbf{(10)} &0.440\textbf{(5)} &0.351\textbf{(5)} &2.385\textbf{(15)} &5.01\textbf{(30)} \\ 
 41&220460648 & 17.193&1.715169(6) &0.583033(4) &-2.05\textbf{(7)}&0.2826\textbf{(16)} &0.148\textbf{(8)} &0.825\textbf{(29)} &0.474\textbf{(7)} &0.351\textbf{(6)} &2.397\textbf{(17)} &5.11\textbf{(33)} \\ 
 42&220489863 & 17.700&1.672909(5) &0.597761(3) &-1.70\textbf{(7)} &0.3223\textbf{(17)} &0.147\textbf{(4)} &0.955\textbf{(22)} &0.498\textbf{(6)} &0.355\textbf{(6)} &2.406\textbf{(16)} &5.15\textbf{(32)} \\ 
 43&220528452 & 17.651&1.795390(5) &0.556982(3) &-1.10\textbf{(7)} &0.2955\textbf{(12)} &0.152\textbf{(8)} &0.843\textbf{(32)} &0.493\textbf{(5)} &0.353\textbf{(5)} &2.423\textbf{(14)} &5.20\textbf{(30)} \\ 
 44&228881816 & 15.279&1.742731(6) &0.573812(3) &-1.05\textbf{(5)} &0.2881\textbf{(11)} &0.167\textbf{(7)} &0.847\textbf{(20)} &0.498\textbf{(5)} &0.333\textbf{(5)} &2.515\textbf{(13)}&5.37\textbf{(29)} \\ 
 45&228930895 & 16.177&1.711491(7) &0.584286(4) &-1.75\textbf{(5)} &0.2651\textbf{(14)} &0.158\textbf{(3)} &0.769\textbf{(16)} &0.462\textbf{(6)} &0.350\textbf{(6)} &2.405\textbf{(16)} &5.12\textbf{(31)} \\ 
 46&229105036 & 14.059&1.488529(5) &0.671804(3) &-1.60\textbf{(6)} &0.3249\textbf{(14)} &0.167\textbf{(2)} &0.954\textbf{(22)} &0.530\textbf{(5)} &0.339\textbf{(5)} &2.681\textbf{(13)} &5.47\textbf{(29)} \\ 
 47&246785473 & 13.367&1.631931(4) &0.612771(2) &-1.25\textbf{(7)} &0.3416\textbf{(4)} &0.202\textbf{(4)} &0.989\textbf{(12)} &0.452\textbf{(3)} &0.272\textbf{(3)} &2.697\textbf{(8)} &5.70\textbf{(22)} \\ 
 48&246872772 & 17.710&1.738429(7) &0.575232(4) &-1.25\textbf{(5)} &0.2197\textbf{(9)} &0.165\textbf{(3)} &0.608\textbf{(15)} &0.447\textbf{(4)} &0.333\textbf{(4)} &2.388\textbf{(12)} &5.05\textbf{(27)} \\ 
 49&248419289 & 15.718&1.606464(4) &0.624485(3) &-1.35\textbf{(6)} &0.3388\textbf{(4)} &0.227\textbf{(3)} &1.006\textbf{(9)} &0.418\textbf{(3)} &0.231\textbf{(2)} &2.688\textbf{(7)} &5.71\textbf{(21)} \\ 
 50&248419294 & 17.372&1.828708(6) &0.546834(4) &-1.60\textbf{(6)} &0.3537\textbf{(12)} &0.148\textbf{(12)} &1.050\textbf{(18)} &0.487\textbf{(7)} &0.363\textbf{(7)} &2.344\textbf{(18)} &4.99\textbf{(34) }\\ 
 51&248503821 & 16.175&1.649460(5) &0.606259(3) &-1.35\textbf{(7)} &0.3237\textbf{(12)}&0.180\textbf{(5)} &0.719\textbf{(25)} &0.489\textbf{(4)} &0.318\textbf{(4)} &2.578\textbf{(11)} &5.35\textbf{(27)} \\ 
 52&248529108 & 16.124&1.713629(6) &0.583557(4) &-1.40\textbf{(6)} &0.2377\textbf{(8)} &0.182\textbf{(7)} &0.674\textbf{(12)} &0.503\textbf{(4)} &0.308\textbf{(3)} &2.638\textbf{(9)} &5.58\textbf{(25)} \\
 53&248561687 & 17.956&1.825950(5) &0.547660(2) &-2.80\textbf{(6)} &0.3176\textbf{(15)} &0.147\textbf{(5)} &0.943\textbf{(20)} &0.498\textbf{(5)} &0.348\textbf{(5)} &2.404\textbf{(14)} &5.15\textbf{(30)} \\ 
 54&248565314 & 16.558&1.998968(4) &0.500258(2) &-1.95\textbf{(5)} &0.3503\textbf{(17)} &0.141\textbf{(4)} &1.030\textbf{(34)} &0.485\textbf{(6)} &0.359\textbf{(5)} &2.344\textbf{(15)} &5.02\textbf{(30)} \\ 
 55&248591089 & 15.554&1.873048(5) &0.533889(2) &-1.30\textbf{(7)} &0.3181\textbf{(15)} &0.148\textbf{(1)} &0.952\textbf{(19)} &0.507\textbf{(6)} &0.357\textbf{(5)} &2.404\textbf{(14)} &5.16\textbf{(30)} \\ 
 56&248828350 & 14.476&1.519119(5) &0.658276(4) &-2.25\textbf{(6)} &0.2925\textbf{(14)} &0.152\textbf{(5)} &0.860\textbf{(20)} &0.479\textbf{(5)} &0.355\textbf{(5)} &2.421\textbf{(14)} &5.16\textbf{(29)} \\ 
 57&251316614 & 15.626&2.197720(4) &0.455017(1) &-1.30\textbf{(6)} &0.4065\textbf{(19)} &0.138\textbf{(10)} &1.142\textbf{(20)} &0.481\textbf{(5)} &0.346\textbf{(5)} &2.279\textbf{(14)} &4.83\textbf{(30)} \\ 
\enddata
\end{deluxetable}
\end{longrotatetable}

\section{Analysis Results} 
\label{sec:results}
\subsection{The properties of Fourier composition coefficients}
\begin{figure}
\begin{center}
\includegraphics[width=3.5 in]{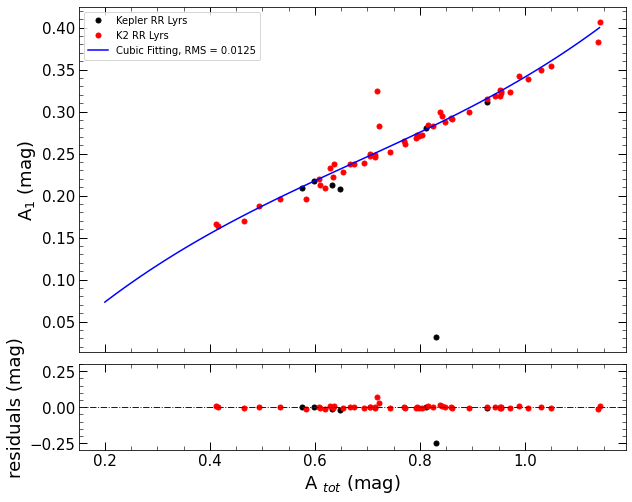}
\caption{The relationship between A$_{tot}$ and A$_1$. The black and red dots are the RRLs in this study observed by $\sl Kepler$ and K2, respectively. The blue curve of the data points as shown in the top panel, and the bottom panel shows the residuals.}\label{fig:Atot}
\end{center}
\end{figure}

\begin{figure}
\begin{center}
\includegraphics[width=3.5 in]{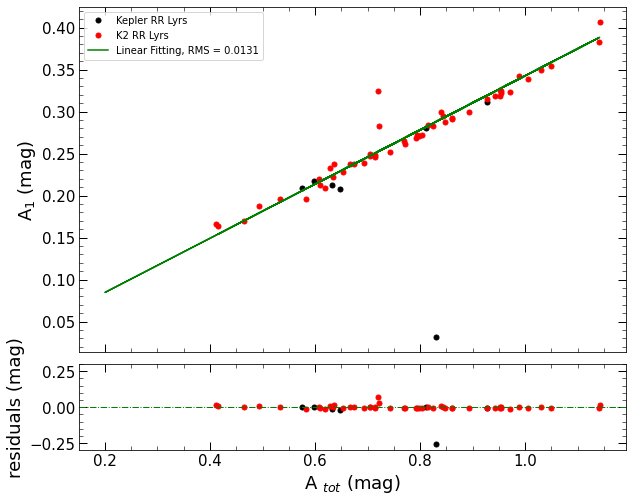}
\caption{The relationship between A$_{tot}$ and A$_1$. The black and red dots are the RRLs in this study observed by $\sl Kepler$ and K2, respectively. The green line is the linear fitting of the data points as shown in the top panel, and the bottom panel shows the residuals.}\label{fig:Linear}
\end{center}
\end{figure}

The Fourier coefficient of $A_1$ is approximately proportional to the total amplitude of A$_{tot}$, since it is the dominant component of A$_{tot}$. 
In order to illustrate this relationship between A$_1$ and A$_{tot}$, we use a cubic and a linear equation to fit the two coefficients A$_1$ and A$_{tot}$ following \cite{2011MNRAS.417.1022N} who found that the dependence between the two coefficients might be cubic using the 19 non-Blazhko RRab stars observed by $\sl Kepler$ and \cite{2014MNRAS.445.1584S} who analysed 176 non-Blazhko RRab stars from ASAS and WASP surveys found that the relation might be linear for the two coefficients, respectively. The cubic fitting is as follows,
\begin{equation}
\centering
\begin{aligned}
A_1 = &-0.28(1)\times A_{tot}^3 -0.56(5)\times A_{tot}^2 \\
&+ 0.66(2)\times A_{tot} -0.004 (1)
\end{aligned}
\label{eq:A1_Atot}
\end{equation}
while the linear fitting is as follows,
\begin{equation}
\centering
\begin{aligned}
A_1 = 0.3223(1) \times A_{tot} + 0.0202 (1)
\end{aligned}
\label{eq:A1_Atot_Linear}
\end{equation}
When the cubic curve (as shown in the top panel of Figure~\ref{fig:Atot}) is subtracted from the A$_{tot}$-A$_1$  diagram, the residuals show no significant deviations from the average value of zero with the rms of 0.0125 mag (as shown in the bottom panel of Figure~\ref{fig:Atot}). But for the linear fitting, although the residuals show no significant variations (presented in the bottom panel of Figure~\ref{fig:Linear}) after subtracting the linear trending (presented in the top panel of Figure~\ref{fig:Linear}) from the A$_{tot}$-A$_1$ diagram, the value of rms is 0.013 mag, which is slightly larger than that of the cubic fitting. It is worth mentioning that the data points away from 3$\sigma$ are not considered for those two kinds of fittings as shown in the top panels of the two figures.


\begin{figure}
\includegraphics[width=3.5 in]{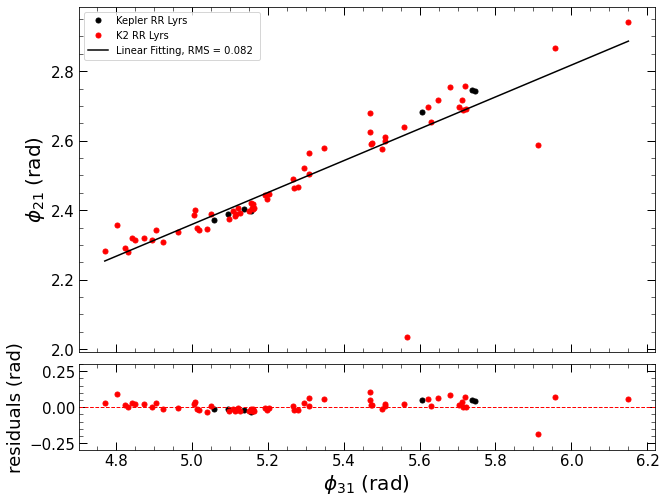}
\caption{The relationship of $\phi_{31}$ and $\phi_{21}$. The black and red dots represent RRLs in this study in the $\sl Kepler$ and $\sl K2$ fields, respectively. The dark line is the linear equation between the two coefficients with a rms of 0.082 rad as shown in the top panel. After extracting this fit, the residuals are shown in the bottom panel.}\label{fig:phi}
\end{figure}

\cite{2014MNRAS.445.1584S} and \cite{2011MNRAS.417.1022N} had investigated the relation between the two Fourier coefficients $\phi_{21}$ and $\phi_{31}$ of the non-Blazhko RRab stars, and they both pointed out that the two coefficients of the stars follow a linear relation. In this work, we perform a linear fitting for the two Fourier coeﬀicients $\phi_{21}$ and $\phi_{31}$ and the fitting equation is as following,
\begin{equation}
\centering
\phi_{21} = 0.459 (10) \times \phi_{31} - 0.064 (28) 
\label{eq:linear}
\end{equation}
The fitting is shown in the top panel of Figure~\ref{fig:phi}, and the residuals are plotted in the bottom panel of the figure, with an rms of 0.082 rad. The data points beyond 3$\sigma$ away from the trending are not considered in the fitting.

\begin{figure*}
\includegraphics[width=8.0 in]{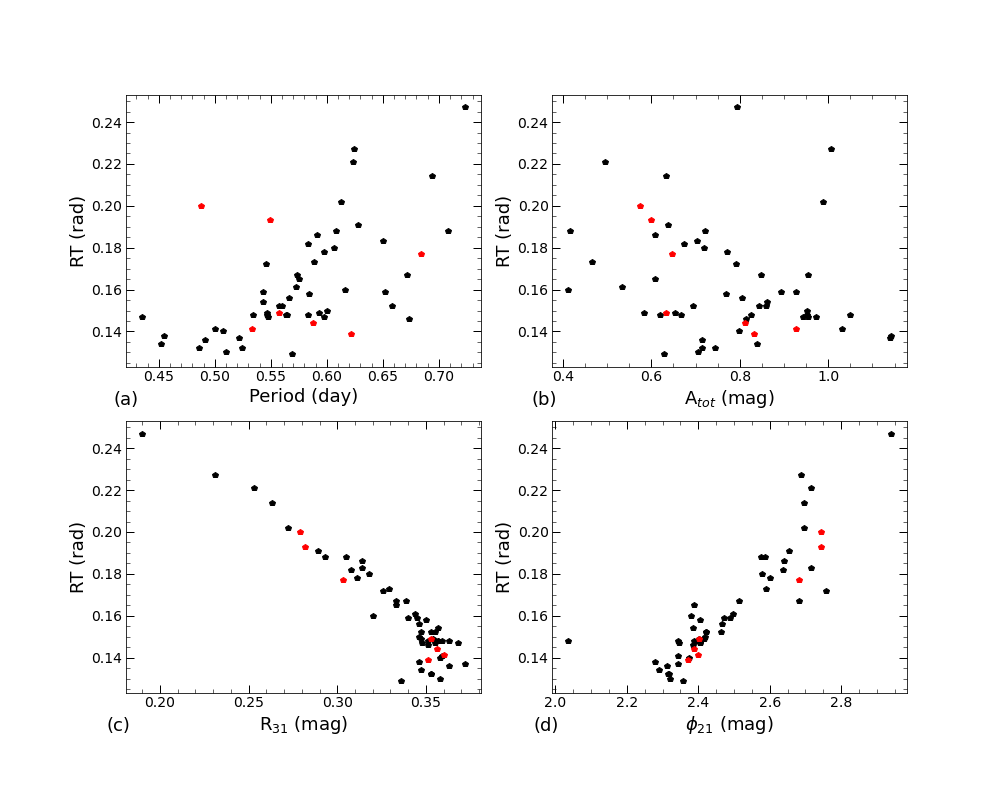}
\caption{Rise time as a function of the parameters of the light curves. The non-Blazhko RRab stars in this study observed by $\sl Kepler$ and $\sl K2$ are presented in red and black dots, respectively.}
\label{fig:RT}
\end{figure*}

Figure~\ref{fig:RT} shows how the rise time (RT) correlates with the parameters of the light curves, including the fundamental period, total amplitude $A_{tot}$, the amplitude ratio $R_{31}$ and the phase difference $\phi_{21}$. Panel (a) shows the relation between RT and the main pulsation period. It can be seen that the longer period probably corresponds to a higher rise time for most of the stars. This is for the first time that this tendency is noticed for $\sl Kepler$ and $\sl K2$ non-Blazhko RRab stars. The data points of the total amplitudes A$_{tot}$ versus RT are scattered as shown in panel (b). R$_{31}$ and RT might follow a roughly linear trend shown in panel (c). The distribution of RT versus $\phi_{21}$ is scattered as shown in panel (d).

\subsection{Period-$\phi_{31}$-[Fe/H]}
\begin{figure}
\centering
\includegraphics[width=3.5 in]{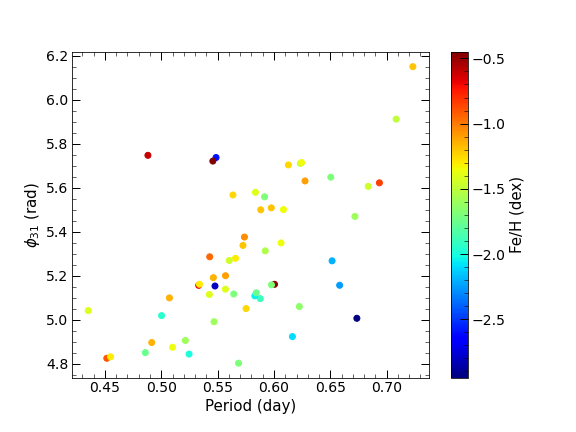}
\caption{Period-$\phi_{31}$-[Fe/H] fit for the RRLs observed in the LAMOST-$Kepler$/$K2$ project. The color bar is the value of the period.}
\label{fig:PHI31}
\end{figure}

A series studies \citep{1996A&A...312..111J,1996ApJ...466L..17K, 2001A&A...371..579K, 2009MNRAS.397..350J,2013ApJ...773..181N,2016CoKon.105...19P,2016ApJS..227...30N, 2021MNRAS.502.5686I,2021ApJ...912..144M} have been conducted to investigate the relation of Period-$\phi_{31}$-[Fe/H]. In this study, we use the fundamental periods P and the phase differences $\phi_{31}$ of the 54 non-Blazhko RRab stars with the mental abundances [Fe/H] of these stars provided by the LAMOST-$\sl Kepler$/$\sl K2$ project to give a new calibration of the relation. This relation could be fitted by following equation,
\begin{equation}
\centering
{\rm [Fe/H]} = a + b\times(P- \overline P_0) + c\times(\phi_{31}-\overline{\phi}_{31})
\label{eq:FEH2}
\end{equation}
where $\overline P_0$ and $\overline{\phi}_{31}$ are the mean values of the fundamental periods and phase differences $\phi_{31}$ of the 54 non-Blazhko stars studied in this work, respectively. We adopt the least squares method which is implemented by SciPy package \citep{2020NatMe..17..261V} to estimate the best fitting coefficients with their corresponding standard errors, for which, a = -3.650 (1), b = 0.848 (4) and c = -3.992 (1). Figure~\ref{fig:PHI31} illustrates the distribution of the [Fe/H] in the $\phi_{31}$ versus periods plane of those RRLs.

\section{Discussion}
\subsection{The properties of Fourier composition coefficients}
In this study, we find that the values of standard errors of cubic fit and linear fit for the relation of the amplitude of the primary frequency A$_{1}$ and the total amplitude A$_{tot}$ of the 54 non-Blazhko RRab stars are $\sigma_{1}$ = 0.013 and $\sigma_{2}$ = 0.0125, respectively, which means that they follow either a cubic or linear relation. We do not find any relation for RT with the fundamental period, total amplitude A$_{tot}$ and phase difference $\phi_{21}$ of the stars studied in this work. However, \cite{2014MNRAS.445.1584S} suggested that the RT follows a linear relation with those parameters for the non-Blazhko RRab stars but does not for the Blazhko RRab stars, which is not consistent with our result. This might be due to that the precision of the photometry from $Kepler$ and $K2$ is much higher than that of the photometric data in the previous study. As far as the relation between $\phi_{21}$ and $\phi_{31}$ of the stars, we find that they follow a linear relation. But \cite{2014MNRAS.445.1584S} found that the dependence of the two coefficients of $\phi_{21}$ and $\phi_{31}$ is not very strong.

\subsection{Comparing with RRab stars in globular clusters and LMC field}
\begin{figure*}
\begin{center}
\includegraphics[width=7.0in]{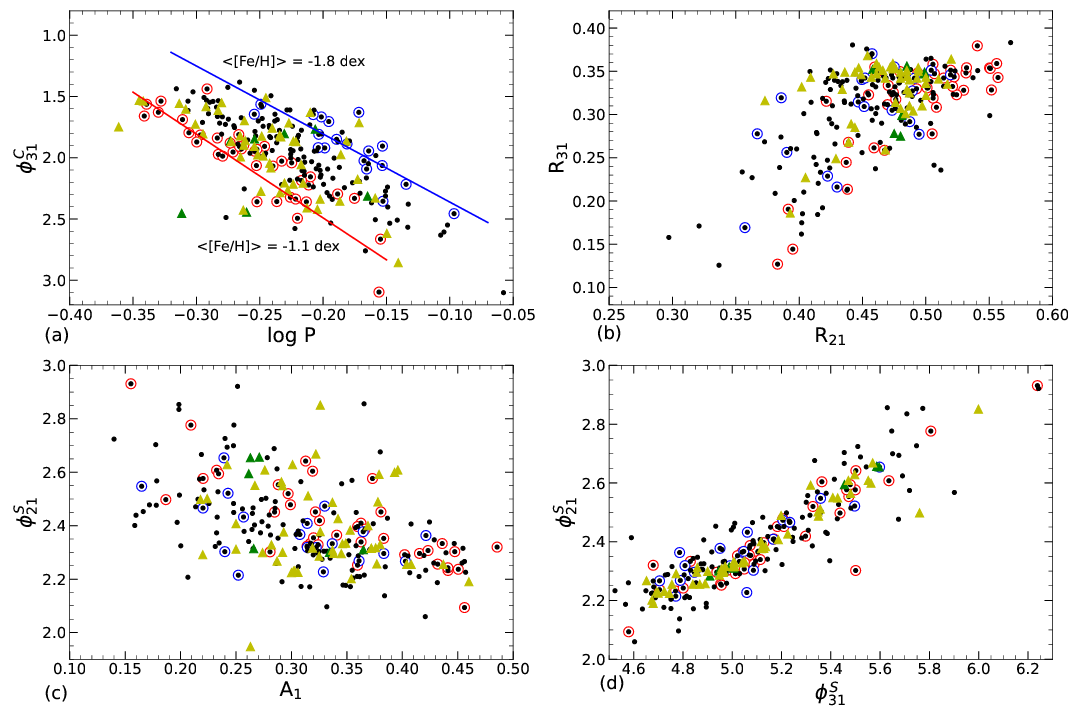}
\caption{Correlation of properties of LAMOST-$\sl Kepler$/$\sl K2$ non-Blazhko RRab stars with the 177 RRab stars located in several Galactic and LMC GCs (black small dots ). The cluster RRab stars presented here are taken from \cite{2001A&A...371..579K} which were adopted by \cite{2011MNRAS.417.1022N}. The green and yellow triangles are the stars in $\sl Kepler$ and $\sl K2$ fields in this study, respectively. The 19 most metal-poor stars, whose metallicities are between -1.70 and -1.99 dex with a mean value of -1.80 dex, are marked by blue circles. The 39 most metal-rich stars, whose metallicities are in the range of -0.97 to -1.23 dex with a mean value of -1.10 dex, are marked by red circles. The blue and red lines in panel (a) are linear fitting for the metal-poor and intermediate metal abundance stars, respectively. Note that the superscript 'S' and 'C' of the coefficients signify phase-parameters computed with sine and cosine series, respectively.}
\label{fig:GC}
\end{center}
\end{figure*}

As the Fourier decomposition coefficients derived from light curves observed in $Kepler$ white band of the 54 non-Blazhko RRab, we convert them into $\sl V$ mag using formula (2) of \cite{2011MNRAS.417.1022N}. Figure~\ref{fig:GC} shows the correlations of properties of the non-Blazhko RRab stars in this study with the 177 RRab stars located in several Galactic and LMC GCs \citep{2001A&A...371..579K}. Panel (a) shows the logP-$\phi_{31}$ diagram. The stars in the GCs with poor and intermediate metallicities define clearly two edges, respectively. The metal-poor subsample consists of 19 stars, whose metallicities are in the range from -1.70 to -1.99 dex with an average value of -1.8 dex \citep{2001A&A...371..579K,2011MNRAS.417.1022N}. The other subsample stars are the intermediate-metallicity stars containing 39 members, whose metallicities are between -0.97 and -1.23 dex with an average value of -1.1 dex. It is clear that most no-Blazhko RRab stars in LAMOST-$Kepler$/$K2$ fields have metal abundance between -1.80 dex and -1.10 dex, except for two $Kepler$ stars and five $K2$ stars with metallicities higher than -1.1 dex. Panel (b) shows that the distribution of the Fourier coefficients R$_{21}$ and R$_{31}$ of the 54 non-Blazhko RRab stars in the $\sl Kepler$/$\sl K2$ survey are similar to the RRab stars in GCs. For the stars studied in this work, we find that some of them with low $R_{31}$ values have relatively high $R_{21}$ values. We also find that most stars are in the upper right side of panel (b) and most of them have high metallicities. Panel (c) shows that the stars studied in this work do not differ from the stars in GCs. We also note that the metallcicites might have no significant effect in this panel. Panel (d) shows the agreement between the phase parameters of the 54 non-Blazhko RRab stars and globular cluster RRab stars, which supports that the phase parameters of $\phi_{21}^s$ and $\phi_{31}^s$ might follow a linear relation. However, panel (d) also presents very little dependence on the metallicities.

As the cluster RRLs adopted in this work cover roughly one dex in the metal intermediate regime, we collected the data of the reference stars from a large sample high resolution spectroscopic surveys of RRLs \citep{2011ApJS..194...38F,2014ApJ...782...59G,2013ApJ...773..181N,2017ApJ...848...68S,2017ApJ...835..187C,2019ApJ...881..104M,2021ApJ...908...20C,2021MNRAS.503.4719G}, with the metallicities of RRLs ranging from -3.0 dex to solar or super-solar iron abundance based on those high resolution spectra. We cross match the catalog of those studies with our data for the reference stars. We derived 7 common stars from the study of \cite{2013ApJ...773..181N} but no common stars from other literature \citep{2011ApJS..194...38F,2014ApJ...782...59G,2017ApJ...848...68S,2017ApJ...835..187C,2019ApJ...881..104M,2021ApJ...908...20C,2021MNRAS.503.4719G} and those stars are all in the $\sl Kepler$ field. But we notice that our database of metallicity of RRLs studied in this work is a subsample of \cite{2020ApJS..247...68L}, who collected the data of the reference stars with reliable metallicity estimates either from high-resolution spectroscopy with the metallicity ranging from -2.95 dex to -0.59 dex \citep{1995AJ....110.2319C,2011ApJS..194...38F,2012MNRAS.422.2116K,2013ApJ...773..181N,2014ApJ...782...59G,2015MNRAS.447.2404P} or as a member star of  globular clusters \citep{2010arXiv1012.3224H} with the metallicity range from -2.37 dex to -1.29 dex. They finally obtained 47 stars in common, which formed their reference star sample. The metallicity scale adopted by them was the one established by \cite{2009A&A...508..695C}, which was derived from the old metallicity scale \citep{1984ApJS...55...45Z}. They found that the values of metallicities of RRLs estimated from the low resolution spectra of LAMOST DR6 agree well with those of the compiled reference stars, with a negligible offset of -0.04 dex and a standard deviation of 0.22 dex. They also found that the dispersion is comparable to that yielded by multi-epoch observations.

\begin{figure*}
\begin{center}
\includegraphics[width=7.0 in]{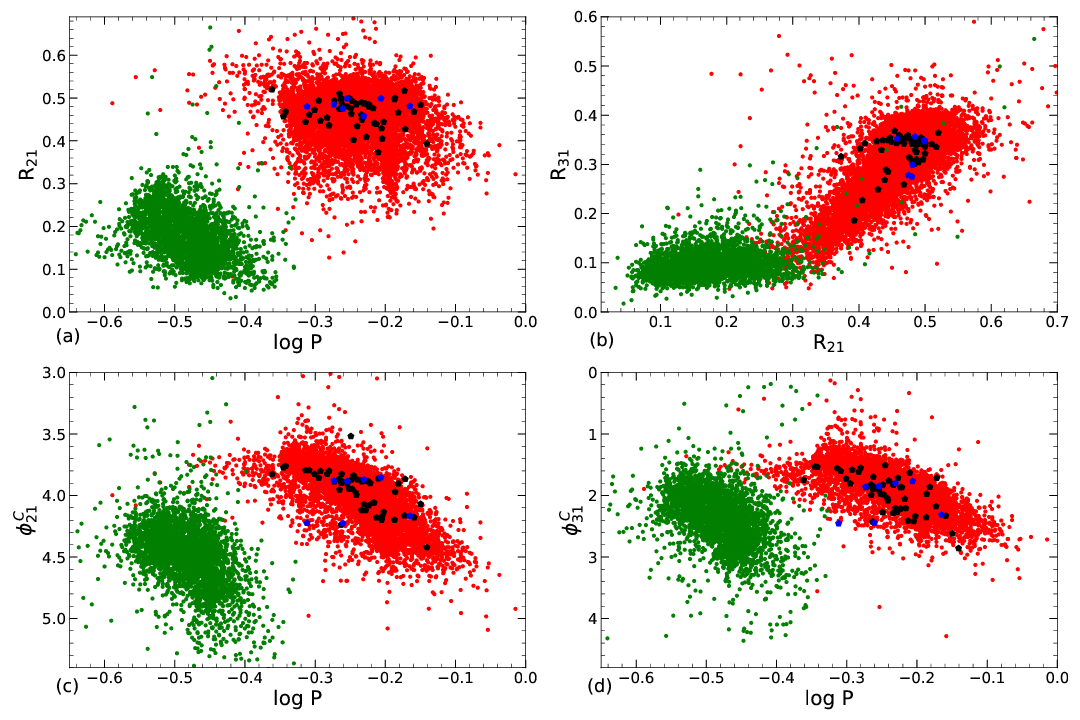}
\caption{Comparison of Fourier coefficients of the non-Blazhko RRab stars in this study with those derived from the OGLE-IV LMC field RR Lyrae stars provided by \cite{2016AcA....66..131S} and the known non-Blazhko RRab stars observed by $\sl Kepler$ determined by \cite{2011MNRAS.417.1022N}. The red points represent RRab, green points RRc stars. The blue pentagon are stars observed by both $\sl Kepler$ and LAMOST DR6. The dark pentagon are the stars observed by both $\sl K2$ and LAMOST DR6.}
\label{fg:Com_LMC}
\end{center}
\end{figure*}

We also compare the Fourier decomposition coefficients R$_{21}$, R$_{31}$, $\phi_{21}^c$ and $\phi_{31}^c$ with those derived for the RRLs in the central regions of the LMC. The latter are determined from the OGLE Collection of Variable Stars by \cite{2016AcA....66..131S} and transformed from $\sl I$ to $\sl V$ band using equations provided by \cite{1998AcA....48..341M}. It is clear that the coefficients of the 54 non-Balzhko RRab stars determined in this work agree with the coefficients of the RRab stars well and differs from other type RRLs shown in Figure~\ref{fg:Com_LMC}.

\subsection{Period-$\phi_{31}$-[Fe/H]}
\begin{figure*}
\centering
\includegraphics[width=7.7in]{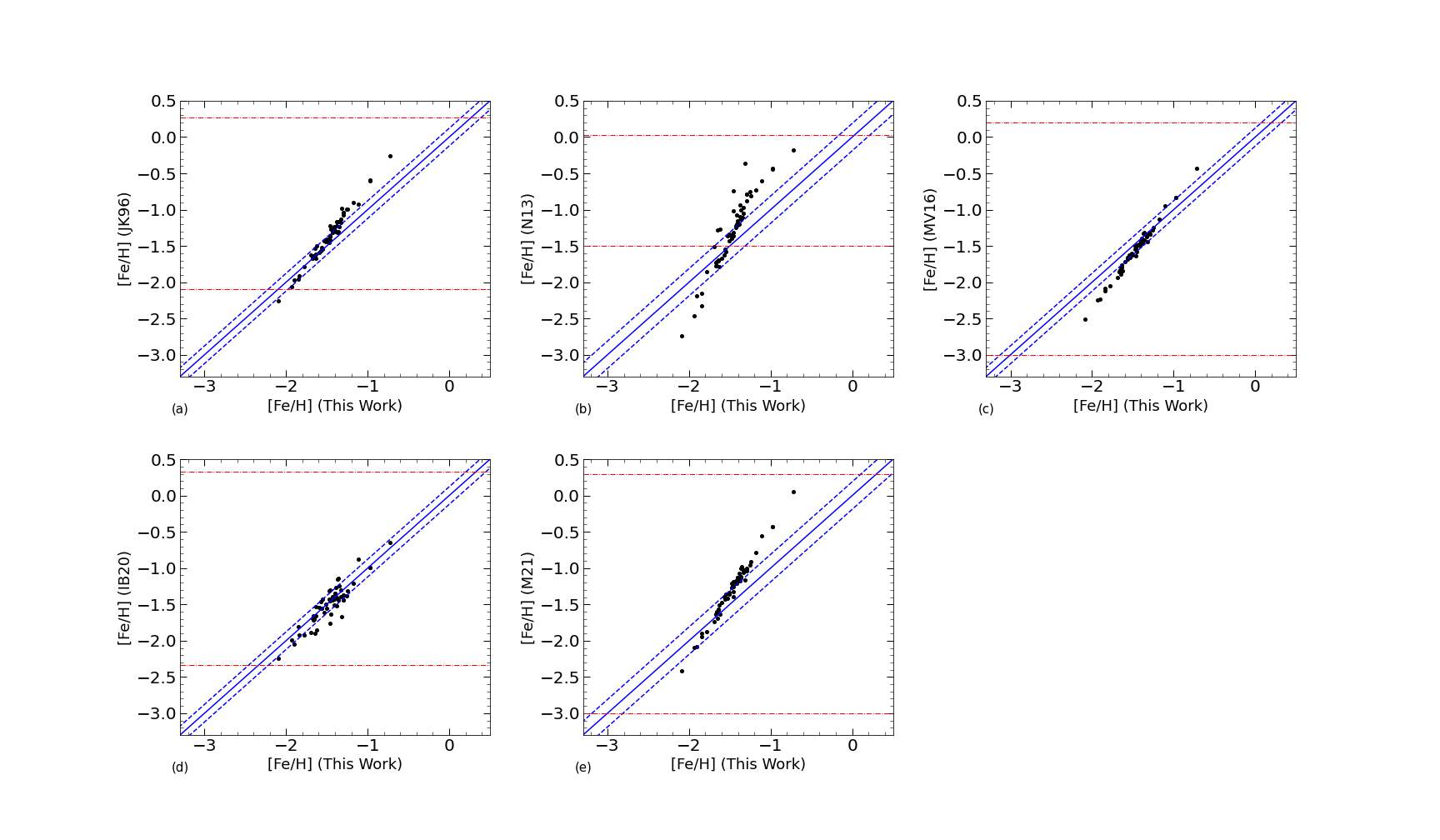}
\caption{Comparison of the [Fe/H] values of the non-Blazhko RRab stars in Table~\ref{tab:parameters} derived by Eq~\ref{eq:FEH2} with those obtained from the relationship of \cite{1996A&A...312..111J}, \cite{2013ApJ...773..181N}, \cite{2016CoKon.105...53M}, \cite{2021MNRAS.502.5686I} and \cite{2021ApJ...912..144M}, respectively. The blue solid lines are 1:1 secants, and the blue dashed lines are the 1 $\sigma$ dispersions of those differences of the metallicity derived by the equation in previous literature and ours. The range of calibrated metallicities used in each work to derive their respective relations are noted in each panel with red dashed–dotted lines.}
\label{fig:Relation}
\end{figure*}

We compare the metallicities [Fe/H] of the stars studied in this wrok calculated using Eq.~\ref{eq:FEH2} with those derived by relationships as documented in the literature \citep{1996A&A...312..111J,2013ApJ...773..181N,2016CoKon.105...53M,2021MNRAS.502.5686I,2021ApJ...912..144M}. For consistency, the metallicities of those investigations are converted to the often-used scale of \cite{2009A&A...508..695C} (hereafter C09). The result of this comparison is shown in Figure~\ref{fig:Relation}, the abscissa of all subgraphs refer to the metallicities of the stars calculated using Eq.~\ref{eq:FEH2}, the ordinate of all subgraphs refer to the metallicities of the stars derived with the relations in the literature. Panel (a) shows the metallicities comparison between ours and those of \cite{1996A&A...312..111J} (hereafter JK96), whose relation was derived using photometric data of 81 field RRab stars in $V$ band with metallicities based on the high-dispersion spectroscopy scale of \cite{1995AcA....45..653J}. For consistency, we first convert the metallicities derived with their relation to the C09 scale using the formula provided by \cite{2011MNRAS.414.1111K}: [Fe/H]$_{C09}$ = 1.001[Fe/H]$_{JK96}$-0.112. We then convert the Fourier coefficient $\phi_{31}$ in K$_p$ system to V-band with the formula (2) of \cite{2011MNRAS.417.1022N}. We find that the comparing result between the two relations is different obviously within the calibration rang (-2.1 dex $\leq$ [Fe/H] $\leq$ 0.7 dex) of JK96 (red horizontal lines) particularly in the metal-rich regime. This might be due to that the photometric data of JK96 were collected from heterogeneous observations at various sites and were either lack of phase coverage or had excessive noise which caused a failure of Fourier fit in JK96. 
\cite{2013ApJ...773..181N} (hereafter N13) derived a quadratic Period-$\phi_{31}$-[Fe/H] relation using 19 RRab stars in the $\sl Kepler$ field with accurate metallicities measurements. The metallicities comparison between ours and those of N13 is shown in panel (b). The Fourier coefficients $\phi_{31}$ of the 57 stars are in same $K_p$ system with that of \cite{2013ApJ...773..181N}. Furthermore, the metallicity scale adopted by N11 and ours are in same scale of C09. Note that the scatter is obviously large for either the higher metallicity or lower metallicity in their range ( -1.5 dex $\leq$ [Fe/H] $\leq$ 0.03 dex). \cite{2021ApJ...912..144M} suggested that this might be caused by the higher-order term of the relationship given by N13, and they had only one RRab star with [Fe/H] $\leq$ -2.0 dex resulting in the scarcity of calibrators in their sample for low [Fe/H]. \cite{2016CoKon.105...53M} (hereafter MV16) gave a new calibration of Period-$\phi_{31}$-[Fe/H] based on a sample of 381 RRab stars in the GCs and 8 field RRab stars in order to extend the metallicities range of their samples. The metallicities of their sample were on the C09 scale, 
and we also convert the $K_p$ system $\phi_{31}$ value to the V-band system using formula (2) of \cite{2011MNRAS.417.1022N}. The metallicities comparing between ours with those of MV16 show an obvious scatter within the entire calibration range of MV16, 
as presented in panel (c). This phenomenon might be due to that the sample of 381 RRab stars were binned by period, they computed the mean period, $\phi_{31}$ and V-band amplitude, which means that the calibration provided by MV16 was based on the average instead of the individual properties.
\cite{2021MNRAS.502.5686I} (hereafter IB21) derived a G-band period-$\phi_{31}$-[Fe/H] relation based on the light curves of 84 RRab stars in Gaia Dr2 with known spectroscopic metallicities. For this comparison, we first use formula (2) of \cite{2011MNRAS.414.1111K} to convert the $\phi_{31}$ value in $K_p$ system to the V-band system, then convert it to that in the G-band system using formula (6) of \cite{2016A&A...595A.133C}. What's more, an additional $\pi$ offset should be subtracted from $\phi_{31}$ to set the coefficients on the same scale as IB20 as suggested by \cite{2021ApJ...912..144M}.
However, the metallicity abundances adopted by IB20 were on the scale of \cite{1984ApJS...55...45Z}(ZW). We convert the metallicity abundances of IB20 to the C09 scale using the formula of [Fe/H]$_{C09}$ = 1.105[Fe/H]$_{ZW84}$ + 0.160. The metallicities comparison between ours and those of IB21 exhibit a generally agreement within the entire range of metallicity (-2.53 dex $\leq$ [Fe/H] $\leq$ 0.33 dex), with an rms of 0.123 dex, as shown in panel (d). However, \cite{2021ApJ...912..144M} pointed out that the relation given by IB21 tends to overestimate the metallicity at the metal-poor ends and underestimate the metallicity at the metal-rich ends. Panel (e) shows the metallcicities comparison between ours and those of \cite{2021ApJ...912..144M} (hereafter M21), who used 1980 RRab stars with a metallicity range of -3.0 dex $\leq$ [Fe/H] $\leq$ 0.4 dex. We first convert the $\phi_{31}$ value in $K_p$ system to the V-band system using formula (2) of \cite{2011MNRAS.417.1022N}. A small shift of 0.08 dex is considered in converting to the often-used C09 scale of [Fe/H] as suggested by M21. The result exhibits an obvious scatter between the two relations for the entire range of metallicities. This might be due to that the sample of RRab stars in M21 is significantly larger than ours.

\section{Conclusions}
As a consequence of cross-match the target stars of {\sl Kepler} and {\sl K}2 photometry with those of the spectroscopic observations in LAMSOT DR6, we derive a sample of 57 non-Blazhko RRab stars. The pulsation periods of these stars are determined with the Fourier Decomposition method applied on the light curves, including $R_{21}$, $R_{31}$, $\phi_{21}$, $\phi_{31}$, A$_1$ and A$_tot$. We find that the amplitude ratios R$_{21}$, R$_{31}$ and those phase difference $\phi_{21}$ and $\phi_{31}$ are consistent with those determined in Globular Clusters and LMC. There is a linear relationship between the phase differences of $\phi_{21}$ and $\phi_{31}$, which agrees well with those in the literature \citep{2014MNRAS.445.1584S}. In terms of the amplitudes of the stars studied in this work, we suggest that the amplitudes of primary frequencies A$_1$ and the total amplitudes A$_{tot}$ follow either a cubic or linear pattern, which need investigate in the feature. For the rise time $RT$, we do not find its relevance with the fundamental pulsation period, A$_{tot}$ and $\phi_{21}$. However, it might follow a linear relationship with R$_{31}$.

Based on the homogeneous metallicities, we have derived a new calibration formula for the relationship of period-$\phi_{31}$-[Fe/H], which agrees well with those in the previous studies as documented in the literature \citep{1996A&A...312..111J,2013ApJ...773..181N,2016CoKon.105...53M,2021MNRAS.502.5686I,2021ApJ...912..144M}. We foresee a much larger catalog to be coming as LAMOST is ongoing to release spectra both in low-resolution and medium-resolution for targets with {\sl Kepler} and {\sl K}2 photometry \citep{2020RAA....20..167F}. A further and larger catalog which will refine the calibration of such relationships between different pulsation parameters. Those observational results might bring new constraints to the hydrodynamic models constructed for RR~Lyrae stars in general.

\section*{acknowledgements}
We acknowledge the support from the National Natural Science Foundation of China (NSFC) through grants 11833002, 12090040, 12090042, 12273002 and 12203010. W.Z. is supported by the Fundamental Research Funds for the Central Universities. The Guoshoujing Telescope (the Large Sky Area Multi-object Fiber Spectroscopic Telescope, LAMOST) is a National Major Scientific Project built by the Chinese Academy of Sciences. The authors gratefully acknowledge the Kepler team and all who have contributed to making this mission possible.

\software{astropy \citep{2013A&A...558A..33A, 2018zndo...4080996T}, LightKurve \citep{2018zndo...1181929V,2020zndo...1181928B}, Period04 \citep{2005CoAst.146...53L}}



\bibliography{sample631}{}
\bibliographystyle{aasjournal}


\end{document}